\documentclass[aps,preprint,showpacs]{revtex4}
\usepackage{graphicx,amsbsy,amsmath,amsfonts}
\usepackage{graphicx}
\usepackage[dvips]{color}
\begin{document}

\title{
Recoil-induced subradiance in a cold atomic gas}
\author{M.M. Cola, D. Bigerni, and N. Piovella}
\affiliation{Dipartimento di Fisica, Universit\`a degli Studi di
Milano} \affiliation{I.N.F.N. Sezione di Milano, Via Celoria 16,
Milano I-20133,Italy}

\begin{abstract}
Subradiance, i.e. the cooperative inhibition of spontaneous
emission by destructive interatomic interference, can be realized
in a cold atomic sample confined in a ring cavity and lightened by
a two-frequency laser. The atoms, scattering the photons of the
two laser fields into the cavity-mode, recoil and change their
momentum. Under proper conditions the atomic initial momentum
state and the first two momentum recoil states form a three-level
degenerate cascade. A stationary subradiant state is obtained
after that the scattered photons have left the cavity, leaving the
atoms in a coherent superposition of the three collective momentum
states. After a semiclassical description of the process, we
calculate the quantum subradiant state and its Wigner function.
Anti-bunching and quantum correlations between the three atomic
modes of the subradiant state are demonstrated.
\end{abstract}
\date{\today}
\maketitle
\section{Introduction}

Recent experiments with Bose-Einstein Condensates (BEC) driven by
a far off-resonant laser beam have demonstrated collective
Superradiant Rayleigh \cite{MIT:1,MIT:2,Tokio,LENS} and Raman
scattering \cite{Schneble,Yoshi}, sharing strong analogies with
the superradiant emission from excited two-level atoms
\cite{SR:review}. In these experiments an elongated BEC scatters
the pump photons into the end-fire modes along the major
dimensions of the condensate, acquiring a  momentum multiple of
the two-photon recoil momentum $\hbar\vec q$, where $\vec q=\vec
k-\vec k_s$ and $\vec k$ and $\vec k_s$ are the wave vectors of
the pump and the scattered field. Theoretical works have shown
that the Superradiant Rayleigh Scattering relies on the quantum
collective atomic recoil (QCARL) gain mechanism, in which the fast
escape of the emitted radiation from the active medium leads to
the superradiant emission \cite{Gatelli,JoB,SR:CARL}. The quantum
regime of CARL \cite{CARL:1,CARL:2} occurs when the two-photon
recoil frequency $\omega_r=\hbar q^2/2m$ is larger than the gain
bandwidth, such that the recoil frequency shifts the atoms out of
resonance inhibiting further scattering processes. As a
consequence in the QCARL each atom coherently scatters a single
pump photon, changing momentum by $\hbar q$. The process in which
the atoms make a transition between two momentum states ($\vec
p=0$ and $\vec p=\hbar \vec q$) has strong analogies with that of
two-level atoms prepared in the excited state and decaying to the
lower state by spontaneous and stimulated emission. However, the
incoherent spontaneous emission dominating the two-level atomic
decay is absent in the momentum transition, where spontaneous
emission is associated to momentum diffusion due to the scattering
force, which can be made very small if the laser is sufficiently
detuned from the atomic resonance. The absence of Doppler
broadening and the long decoherence time of a BEC allows to
observe superradiance and coherent spontaneous emission much more
easily than from electronic transitions in excited atoms, in which
the decay is dominated by the incoherent spontaneous emission.

Another example of cooperative phenomena from excited two-level
atoms is subradiance, i.e. the cooperative inhibition of
spontaneous emission by a destructive interatomic interference.
This phenomenon, whose existence has been proposed by Dicke (1954)
in the same article predicting superradiance \cite{Dicke}, has
received less consideration than the more popular superradiance,
also due to the difficulty of its experimental observation. In
fact, the only experimental evidence has been done on 1985 by
Pavolini \textit{et al.} \cite{Pav}. Among different schemes of
multi-level systems in which subradiance  was predicted,
Crubellier \textit{et al.}, in a series of theoretical papers
\cite{Cru1,Cru2,Cru3,Cru4}, proposed a three-level degenerate
cascade configuration in which cooperative spontaneous emission is
expected to exhibit new and striking subradiance effects.

\begin{figure}
\includegraphics[width=6cm]{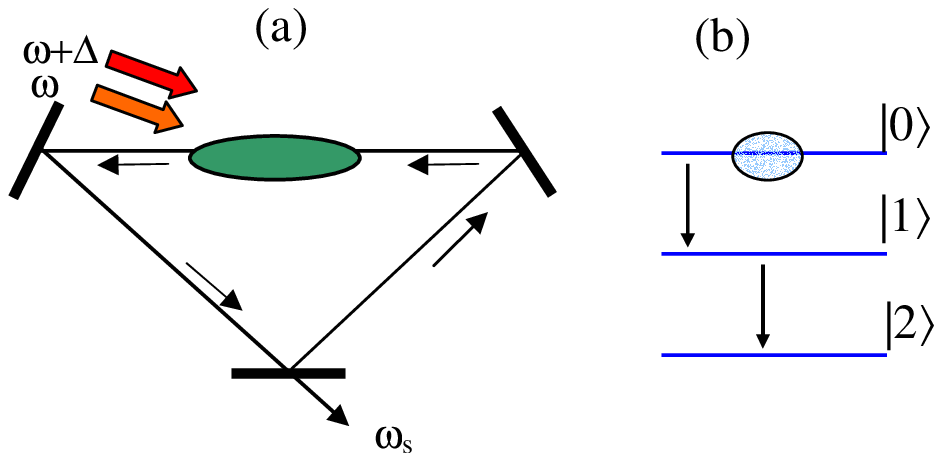}
\caption{(a): Schematic diagram illustrating the geometry of a
two-frequency pump CARL experiment; (b): three-level cascade
scheme.} \label{fig1}
\end{figure}

In this paper we show that subradiance in a three-level degenerate
cascade can be realized in a BEC inserted in a ring cavity and
lightened by two laser fields with frequency difference twice the
two-photon recoil frequency, as illustrated by fig.\ref{fig1}.
The frequency of the scattered photon is determined by energy and
momentum conservation. The process consists in two steps. In the
first step the atoms initially at rest scatter the laser photons
of frequency $\omega$ into the cavity mode of frequency
$\omega_s=\omega-\omega_r$, changing momentum from $0$ to $\vec
p=\hbar\vec q$. In the second step the atom scatters the laser
photon of frequency $\omega+\Delta$ changing their momentum from
$\vec p=\hbar\vec q$ to ${\vec p}\,\,'=2\hbar\vec q$. Since the
change of the kinetic energy of the atom is $\Delta
E=(p\,'\,^2-p^2)/2m=3\hbar\omega_r$, by energy conservation the
frequency of the scattered photon is $\omega+\Delta-3\omega_r$
which coincides with the  frequency generated in the first step
when $\Delta=2\omega_r$. In this way a three-momentum-level
degenerate cascade is realized in which the atoms, initially with
momentum $\vec p=0$, change momentum to the intermediate value
$\hbar\vec q$ and then to the final value $2\hbar\vec q$, emitting
two degenerate photons of frequency $\omega_s=\omega-\omega_r$. In
general the process, as described in ref.\cite{ASR}, will continue
with an other scattering of the photon of frequency
$\omega+\Delta$ changing the atomic momentum from $2\hbar\vec q$
to $3\hbar\vec q$ with the emission of a photon of frequency
$\omega+\Delta-5\omega_r=\omega-3\omega_r$ and so on. However, if
the cavity linewidth is much narrower than the frequency
difference, $\kappa \ll 2\omega_r$, these other frequencies will
be damped out. Then, the oscillation of only the frequency
$\omega_s=\omega-\omega_r$ in the cavity  will restrict the
momentum cascade to the three momentum states, $0$, $\hbar\vec q$
and $2\hbar\vec q$. A basic feature of this system is that the
transition rates are proportional to the pump intensities, so that
they can be varied with continuity. This makes the subradiance
observation much easier than with a three-level cascade between
electronic energy levels, where the transition rates are fixed by
the branching ratios.

\section{Semiclassical treatment}

\subsection{General model}
The quantum collective atomic recoil laser (QCARL) with a
two-frequency pump is described by the following equations for the
order parameter $\Psi(z,t)$ of the matter field  and the cavity
mode field amplitude $a(t)$ \cite{ASR}:
\begin{eqnarray}
i\noindent\frac{\partial\Psi}{\partial t}&=& -\frac{\hbar}{2m}
\frac{\partial^2\Psi}{\partial z^2}+ ig\left[\alpha(t)
a^*e^{i(qz-\delta t)}-{\rm
cc.}\right]\Psi\label{ES}\\
\noindent\frac{da}{dt}&=& gN\alpha(t) \int
dz|\Psi|^2e^{i(qz-\delta t)}-\kappa a, \label{EF}
\end{eqnarray}
where $z$ is the coordinate along the cavity axis and
$\alpha(t)=1+\epsilon\exp(-i\Delta t)$. These equations have been
derived  performing the adiabatic elimination of the atomic
internal degrees of freedom \cite{Gatelli} but replacing the pump
field with $E_p=e^{i(\vec k\cdot\vec x-\omega
t)}\left(E_0+E_1e^{-i\Delta t}\right)$. In Eqs.(\ref{ES}) and
(\ref{EF}) $a(t)=(\epsilon_0 V/2\hbar\omega_s)^{1/2}E_s(t)$ is the
dimensionless electric field amplitude of the scattered radiation
beam with frequency $\omega_s$, $g=(\Omega_0/2\Delta_0)(\omega
d^2/2\hbar\epsilon_0 V)^{1/2}$ is the coupling constant,
$\Omega_0=dE_0/\hbar$ is the Rabi frequency of the pump laser
incident with an angle $\phi$ with respect to the $z$ axis
($\phi=\pi$ if counterpropagating), with electric field $E_0$ and
frequency $\omega$ detuned from the atomic resonance frequency
$\omega_0$ by $\Delta_0=\omega-\omega_0$. The pump laser has a
sideband with frequency $\omega+\Delta$, with $\Delta=2\omega_r$
[where $\omega_r=\hbar q^2/2m$ and $q=2k\sin(\phi/2)$], and
electric field $E_1$ with $\epsilon=E_1/E_0$.  The other
parameters are: $d=\hat\epsilon\cdot\vec d$, the electric dipole
moment of the atom along the polarization direction $\hat\epsilon$
of the laser, $V$, the cavity mode volume, $N$ the total number of
atoms in the condensate, $\delta=\omega-\omega_s$, and $\kappa$,
the cavity linewidth. The emitted frequency $\omega_s$ is within
the cavity frequency linewidth, whereas the pump field is external
to the cavity so that its frequencies are not dependent on the
cavity ones. The order parameter $\Psi$ of the matter field is
normalized such that $\int dz |\Psi|^2=1$.

If the condensate is much longer than the radiation wavelength and
approximately homogeneous, then periodic boundary conditions can
be applied on the atomic sample and the order parameter can be
written as $\Psi(z,t)=\sum_n c_n(t)u_n(z)e^{-in\delta t}$, where
$u_n(z)=(q/2\pi)^{1/2}\exp[in(qz)]$ are the momentum eigenstates
with eigenvalues $p_z=n(\hbar q)$. Using this expansion,
Eqs.(\ref{ES}) and (\ref{EF}) become:
\begin{eqnarray}
\frac{d c_m}{d t}&=&-i\omega_{m}c_m
+g [\alpha(t) a^*c_{m-1}-\alpha^*(t) a c_{m+1}]\label{cm} \\
\frac{da}{dt}& = &g N \alpha(t)\sum_{n}c_{m}c^*_{m+1}-\kappa
a\label{af},
\end{eqnarray}
where $\omega_{n}=n(n\omega_{r}-\delta)$.

\subsection{Three-level approximation}
As as been discussed elsewhere \cite{Gatelli,QCARL}, if the gain
rate is smaller than the recoil frequency the atoms recoil only
along the positive direction of $\vec q$,  absorbing a photon from
the laser and emitting it into the cavity mode. Backward recoil,
in which an atom absorbs a photon from the cavity mode and emits
it into the laser mode, is inhibited by energy conservation. In
this way, the laser photon of frequency $\omega$ induces a
momentum transition from $m=0$ to $m=1$, emitting in the cavity a
photon with frequency $\omega_s=\omega-\omega_r$; the laser photon
of frequency $\omega+\Delta$ [with $\Delta=2\omega_r$] induces a
momentum transition from $m=1$ and $m=2$, emitting an other photon
of the same frequency $\omega_s$. If the cavity linewidth $\kappa$
is smaller less than $2\omega_r$, only the photons with frequency
$\omega_s$ will survive in the cavity. Since further scattering
would generate photons with frequencies $\omega-m\omega_r$, with
$m=3,5,\dots$, which can not oscillate in the cavity, then the
Hilbert space of the atoms is spanned by only the first three
recoil momentum levels, $m=0,1,2$, and Eqs.(\ref{cm}) and
(\ref{af}) reduce to:
\begin{eqnarray}
\frac{d c_0}{d t}&=&-g \alpha^*(t) ac_{1}\label{c0} \\
\frac{d c_1}{d t}&=&i(\delta-\omega_r) c_1
+g [\alpha(t) a^*c_{0}-\alpha^*(t) a c_{2}]\label{c1} \\
\frac{d c_2}{d t}&=&2i(\delta-2\omega_r)c_2
+g\alpha(t) a^*c_{1}\label{c2} \\
\frac{da}{dt}& = &g N \alpha(t)(c_{0}c^*_{1}+c_{1}c^*_{2})-\kappa
a\label{a3}.
\end{eqnarray}
Eqs.(\ref{c0})-(\ref{a3}) contain fast oscillating terms. They can
be eliminated introducing the slowly varying variable $\tilde
c_2=c_2\exp(i\Delta t)$ and approximating
Eqs.(\ref{c0})-(\ref{a3}) neglecting the fast oscillating terms
proportional to $\exp(\pm i\Delta t)$. In this way
Eqs.(\ref{c0})-(\ref{a3}) reduce to:
\begin{eqnarray}
\frac{d c_0}{d t}&=&-g ac_{1}\label{c0:2} \\
\frac{d c_1}{d t}&=&i(\delta-\omega_r) c_1
+g (a^*c_{0}-\epsilon a \tilde c_{2})\label{c1:2} \\
\frac{d\tilde c_2}{d t}&=&2i(\delta-\omega_r)\tilde c_2
+g\epsilon a^*c_{1}\label{c2:2} \\
\frac{da}{dt}& = &g N(c_{0}c^*_{1}+\epsilon c_{1}\tilde
c^*_{2})-\kappa a\label{a3:2}.
\end{eqnarray}
Eqs.(\ref{c0:2})-(\ref{a3:2}) describe the three-level degenerate
cascade of the atoms driven by two laser fields at frequencies
$\omega$ and $\omega+2\omega_r$, respectively, and interacting
with the self-generated cavity mode at the frequency
$\omega_s=\omega-\omega_r$. Notice that the  second transition
rate, from $m=1$ to $m=2$, is proportional to the two pump
amplitude ratio, $\epsilon$.

\subsection{Subradiance in three-level degenerate cascade}

\begin{figure}[b]
\begin{center}
\includegraphics[width=8cm]{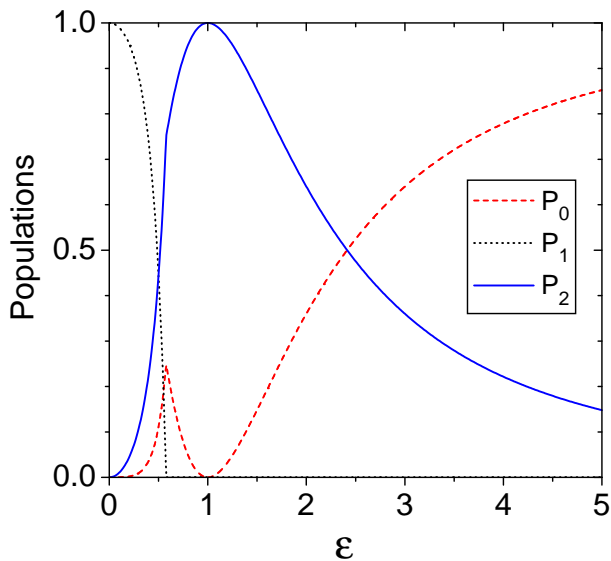}
\end{center}
\vskip 0.2cm \caption{Semiclassical subradiance solution:
populations of the three states, $P_0$ (dashed red line), $P_1$
(dotted gray line) and $P_2$ ( continuous blue line) as a function
of $\epsilon$, as given by Eqs. (\ref{sc:1}) and (\ref{sc:2}).}
\label{fig2}
\end{figure}

Asymptotically, in a time much longer than $1/\kappa$, the photons
leak the cavity and the total polarization in Eq.(\ref{a3:2})
vanishes:
\begin{equation}\label{pol}
    c_{0}c^*_{1}+\epsilon c_{1}\tilde
c^*_{2}=0.
\end{equation}
On resonance ($\delta=\omega_r$) and with the atoms initially at
rest ($c_0(0)=1$), the variables $c_0$, $c_1$, $c_2$ and $a$ are
real and Eqs.(\ref{c0:2})-(\ref{a3:2})  keep invariant the
following quantity:
\begin{equation}\label{inv}
    J=\epsilon^2c_0^2+\tilde c_2^2+2\epsilon c_0\tilde c_2
    =\epsilon^2.
\end{equation}
From it we see that the atoms can not populate completely  the
final state $m=2$ (with $c_2=1$ and $c_0=0$) unless $\epsilon=1$.
Hence, when $\epsilon\neq 1$ the atoms remain in the intermediate
levels $m=0$ and $m=1$ in a subradiant state. Condition
(\ref{pol}), together with the constraints (\ref{inv}) and the
normalization $c_0^2+c_1^2+\tilde c_2^2=1$, determine univocally
the steady-state solution reached asymptotically by the atoms. It
is easy to show that for $\epsilon<1/\sqrt{3}$,
\begin{equation}\label{sc:1}
  c_0=-\epsilon\tilde c_2=\frac{\epsilon^2}{1-\epsilon^2},\quad
  c_1=\left[1-\frac{\epsilon^2(1+\epsilon^2)}{(1-\epsilon^2)^2}\right]^{1/2}
\end{equation}
whereas for $\epsilon>1/\sqrt{3}$,
\begin{equation}\label{sc:2}
  c_0=\frac{1-\epsilon^2}{1+\epsilon^2},\quad
  c_1=0,\quad
  \tilde c_2=-\frac{2\epsilon}{1+\epsilon^2}
\end{equation}
Fig.\ref{fig2} shows the steady-state populations $P_i=|c_i|^2$
for $i=0,1,2$ plotted vs. $\epsilon$. We observe that increasing
$\epsilon$ from zero the population of the intermediate state,
$P_1$, decreases and the population of the final state, $P_2$,
increases. They are equal for $\epsilon=1/2$, with $c_0=1/3$ and
$c_1=-c_2=2/3$. The population of the initial state $P_0$
increases too and reaches a local maximum for
$\epsilon=1/\sqrt{3}$ with $c_0=1/2$, $c_1=0$ and
$c_2=\sqrt{3}/2$. Then, for $\epsilon>1/\sqrt{3}$ the intermediate
state $m=1$ is empty ($c_1=0$) and the population of the initial
state $P_0$ decreases to zero for $1/\sqrt{3}<\epsilon<1$; then
for $\epsilon>1$ it increases until it equals the population of
the final state $P_2$ when $\epsilon=1+\sqrt{2}$. For
$\epsilon>1+\sqrt{2}$ the population of the initial state, $P_0$,
is larger than that of the final state, $P_2$. However, this case
appears stationary only because the semiclassical model neglects
spontaneous emission. The quantum treatment, reported in the next
section, shows that the stationary subradiant state may exist only
for $0<\epsilon<1+\sqrt{2}$.

\begin{figure}[b]
\begin{center}
\includegraphics[width=9 truecm]{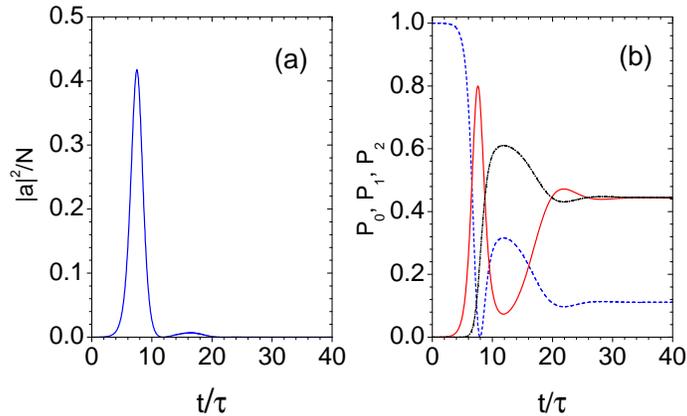}
\end{center}
\vskip 0.2cm \caption{Time evolution of the radiated intensity
$|a|^2$, (a), and the three populations $P_0$ (dashed blue line),
$P_1$ (red continuous line) and $P_2$ (dashed-dotted black line)
vs. $t/\tau$, where $\tau=(g\sqrt{N})^{-1}$, for
$g\sqrt{N}=0.01\omega_r$, $\kappa=0.006\omega_r$,
 $\delta=\omega_r$ and $\epsilon=1/2$.}
\label{fig3}
\end{figure}

\begin{figure}[b]
\begin{center}
\includegraphics[width=8truecm]{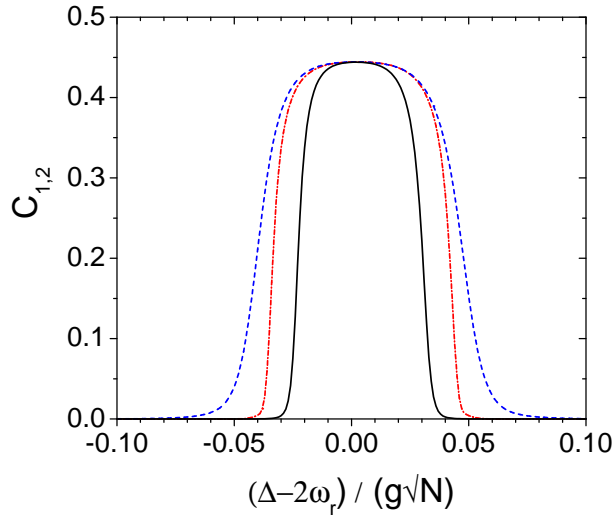}
\end{center}
\vskip 0.2cm \caption{Asymptotic coherence $C_{1,2}=|c_1c_2^*|$
between the intermediate and the final states vs.
$(\Delta-2\omega_r)/g\sqrt{N}$ for $g=0.01\omega_r$,
$\delta=\omega_r$, $\epsilon=1/2$, $\kappa=0.003\omega_r$ (dashed
blue line), $\kappa=0.006\omega_r$ (dashed-dot red line) and
$\kappa=0.012\omega_r$ (continuous black line).} \label{fig4}
\end{figure}

In order to illustrate how the system evolves toward the
subradiance state, fig.\ref{fig3} shows the time evolution of the
field $|a|^2$ (fig.\ref{fig3}a), and the three populations,
(fig.\ref{fig3}b) [$P_0$ (dashed blue line), $P_1$ (red continuous
line) and $P_2$ (dashed-dotted black line)], obtained solving the
complete equations (\ref{cm}) and (\ref{af}) for
$g\sqrt{N}=0.01\omega_r$, $\kappa=0.006\omega_r$,
 $\delta=\omega_r$ and $\epsilon=1/2$, with initial condition
 $c_0(0)=1$, $c_{i\neq 0}(0)=0$ and $a(0)=0.01$. The final populations
 are $P_0=1/9$ and $P_{1}=P_{2}=4/9$, according to Eq. (\ref{sc:1}).
In order to test the dependence of the subradiance state on the
frequency difference $\Delta$ between the two pump fields, figure
\ref{fig4} shows the asymptotic coherence $C_{1,2}=|c_1c_2^*|$
between the intermediate and the final states vs. $\Delta$ for
$g=0.01\omega_r$, $\delta=\omega_r$, $\epsilon=1/2$,
$\kappa=0.003\omega_r$ (dashed blue line), $\kappa=0.006\omega_r$
(dashed-dot red line) and $\kappa=0.012\omega_r$ (continuous black
line). The result shows that subradiance requires a very fine
tuning of the pump frequency difference near $2\omega_r$, within a
precision $\delta\omega\ll g\sqrt{N}$.

\begin{figure}[t]
\begin{center}
\includegraphics[width=9truecm]{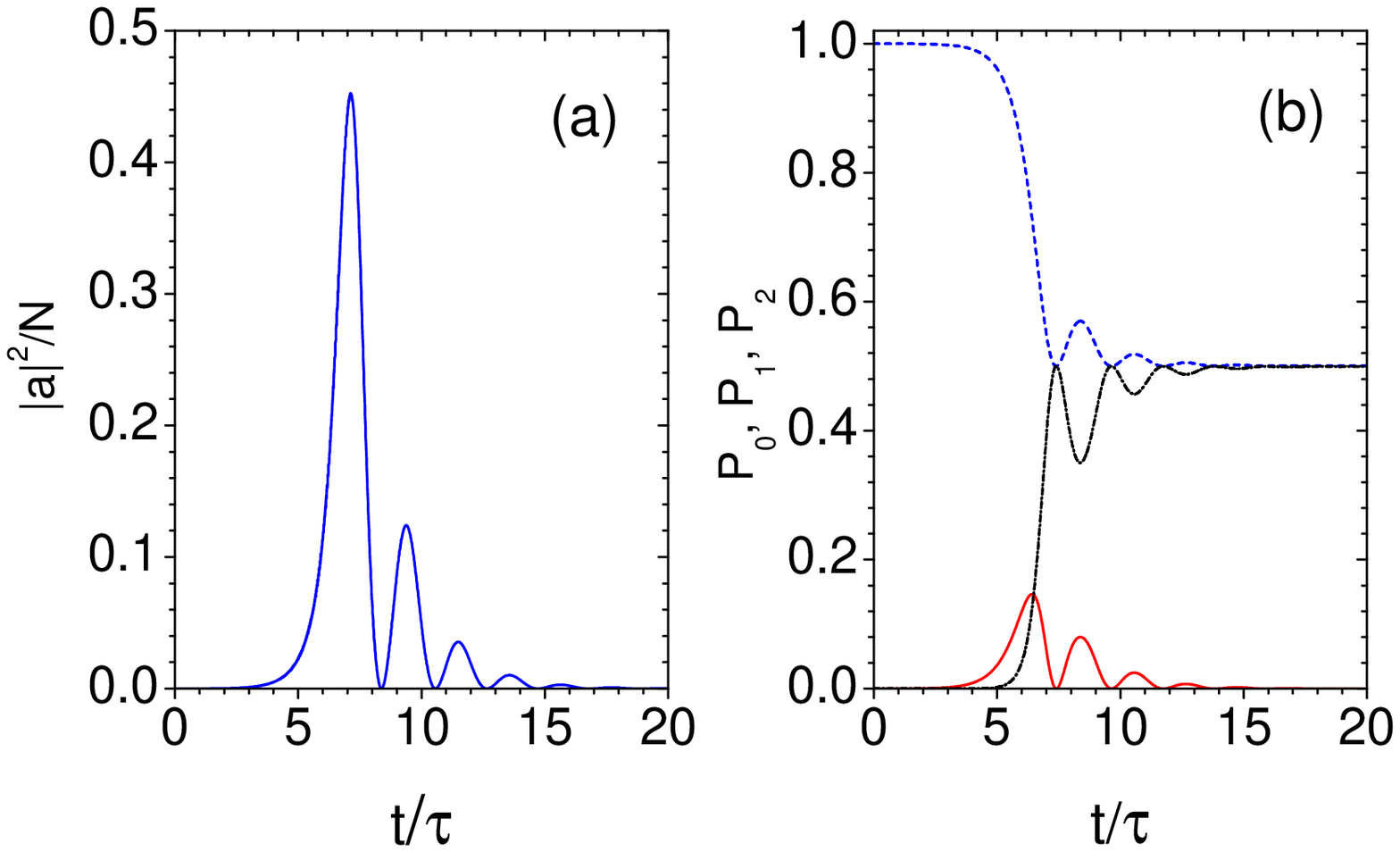}
\end{center}
\vskip 0.2cm \caption{Time evolution of the radiated intensity
$|a|^2$ (a) and the three populations $P_0$ (dashed blue line),
$P_1$ (red continuous line) and $P_2$ (dashed-dotted black line)
as a function of $t/\tau$, where $\tau=(g\sqrt{N})^{-1}$, for the
same parameters of fig.\ref{fig3} and $\epsilon=1+\sqrt{2}$.}
\label{fig5}
\end{figure}

\begin{figure}[t]
\begin{center}
\includegraphics[width=7truecm]{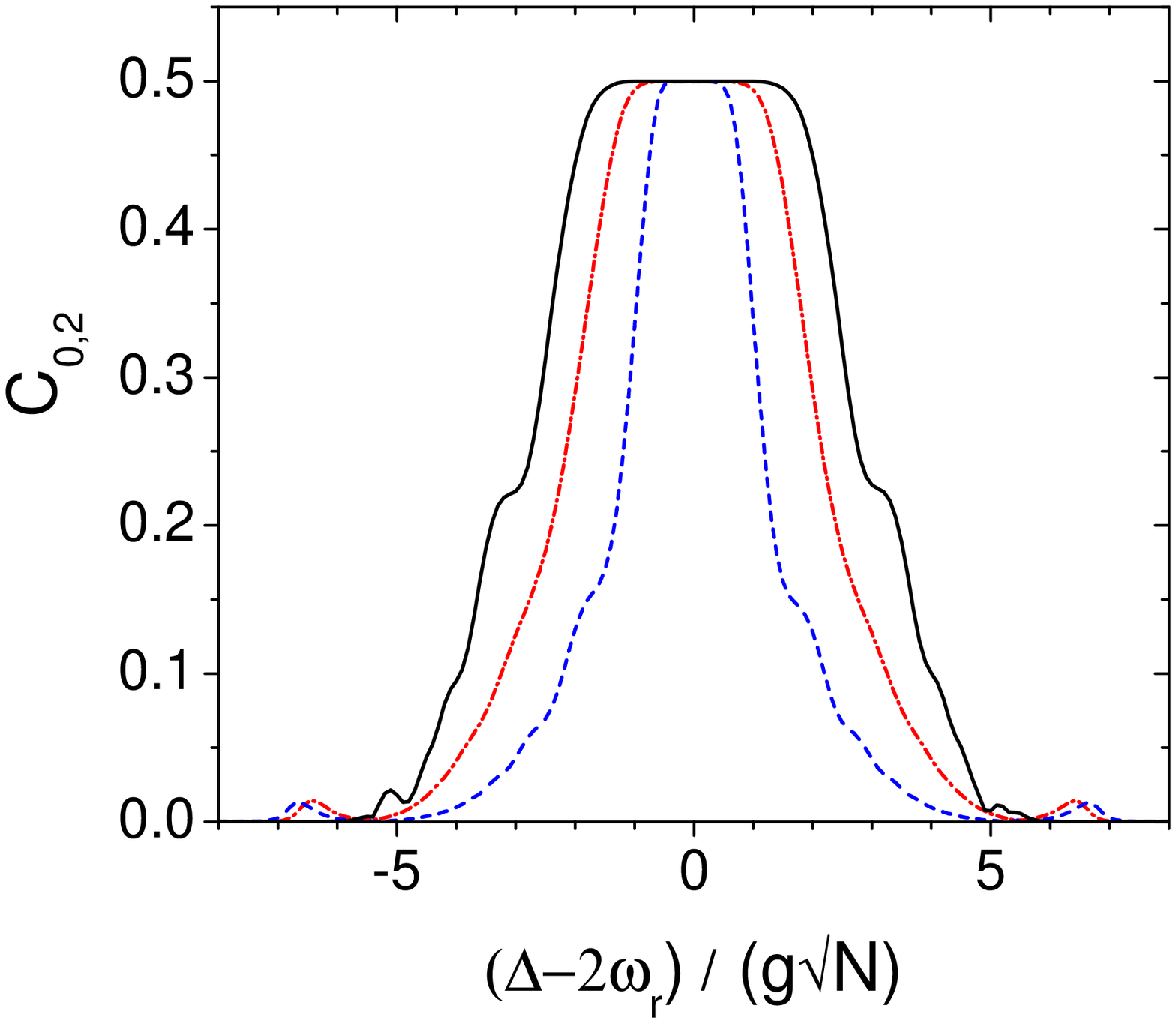}
\end{center}
\vskip 0.2cm \caption{Asymptotic coherence $C_{0,2}=|c_0c_2^*|$
between the initial and the final states vs.
$(\Delta-2\omega_r)/g\sqrt{N}$ for $g=0.01\omega_r$,
$\delta=\omega_r$, $\epsilon=1+\sqrt{2}$, $\kappa=0.003\omega_r$
(dashed blue line), $\kappa=0.006\omega_r$ (dashed-dot red line)
and $\kappa=0.012\omega_r$ (continuous black line).} \label{fig6}
\end{figure}

As a second example, figs. \ref{fig5} and \ref{fig6} show that
same case as in figs.\ref{fig3} and \ref{fig4} but with
$\epsilon=1+\sqrt{2}$. In this case $P_1=0$ and $P_{0}=P_2=1/2$.
We note that whereas in the case $\epsilon=1/2$ the resonance
linewidth of fig.\ref{fig4} decreases when the cavity losses
$\kappa$ increases, on the contrary in the case
$\epsilon=1+\sqrt{2}$ the linewidth increases with $\kappa$ and it
is about a factor $100$ larger. Hence the subradiance with
$\epsilon=1/2$ is more sensible to the frequency mismatch than
that with $\epsilon=1+\sqrt{2}$.

\section{quantum treatment}

\subsection{The subradiant state}
Let now obtain the subradiant state quantum-mechanically, treating
the amplitude $c_n$ as bosonic operators $\hat c_n$ with
commutation rules $[\hat c_m,\hat c_n^\dagger]=\delta_{m,n}$.
Then, according to Eq.(\ref{pol}) the subradiant state
$|sr\rangle$ satisfies:
\begin{equation}\label{sr}
    (\hat c_0\hat c_1^\dagger+\epsilon \hat c_1\hat
    c_2^\dagger)|sr\rangle=0
\end{equation}
(we omit the tilde on $\hat c_2$). It is possible to demonstrate
(see Appendix \ref{App1}) that for a system of $N$ atoms (with $N$
even) there are $N/2$ subradiant states $|sr\rangle_p$, with
$p=1,2,\dots,N/2$, defined as
\begin{eqnarray}\label{state}
    |sr\rangle_p&=&C_p\sum_{k=0}^p\frac{(-1/2\epsilon)^k}{k!}\sqrt{\frac{(2k)!(N-p-k)!}{(p-k)!}}
    \nonumber\\
    &&\times|p-k,2k,N-p-k\rangle,
\end{eqnarray}
where $|m,n,l\rangle=|m\rangle_0 |n\rangle_1 |l\rangle_2$ and
$C_p$ is a normalization constant. The index $p$ is related to the
population difference between the initial and final states, since
$N_0-N_2=2p-N$. The case $p=0$ corresponds to the state
$|0,0,N\rangle$. The link between the subradiant state
$|sr\rangle_p$ and the semiclassical solution (\ref{sc:1}) and
(\ref{sc:2}) is provided by the correspondence between $p$ and the
population difference $N_0-N_2=N(c_0^2-c_2^2)$ in the limit $N\gg
1$. For $\epsilon<1/\sqrt{3}$,
$p=(N/2)(1-2\epsilon^2)/(1-\epsilon^2)$ and for
$\epsilon>1/\sqrt{3}$, $p=N[1-4\epsilon^2/(1+\epsilon^2)^2]$. As
particular cases,  $p=N/3$ for $\epsilon=1/2$  and $p=N/4$ for
$\epsilon=1/\sqrt{3}$. Furthermore, $p=0$ for $\epsilon=1$
 and $p=N/2$ for $\epsilon=1+\sqrt{2}$. Hence $1+\sqrt{2}$ is the
 maximum value of $\epsilon$, giving the
following subradiant state:
\begin{eqnarray}\label{state:N/2}
    |sr\rangle_{N/2}&=&C_{N/2}\sum_{k=0}^{N/2}
    \frac{(-1/2\epsilon)^k}{k!}\sqrt{(2k)!}
    \nonumber\\
    &&\times\left|\frac{N}{2}-k,2k,\frac{N}{2}-k\right\rangle.
\end{eqnarray}
For large $N$, the average value of $k$ is $\langle
k\rangle=1/4\epsilon=(\sqrt{2}-1)/4$, with variance
$\sigma_k^2=\langle k\rangle$.

\begin{figure}[t]
\begin{center}
\includegraphics[width=7cm]{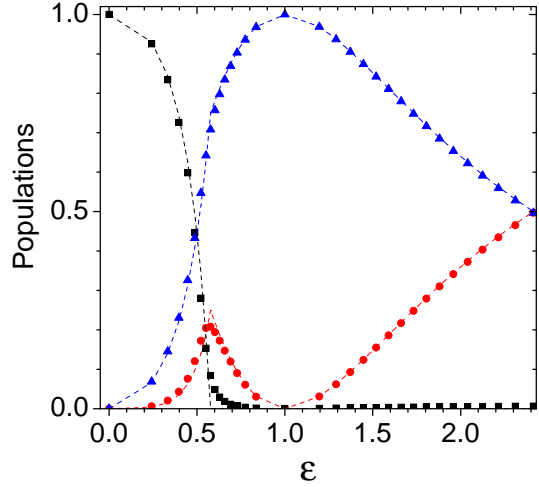}
\end{center}
\vskip 0.2cm \caption{Quantum subradiance solution: population
fraction of the three states, $P_0$ (red circles), $P_1$ (black
squares) and $P_2$ (blue triangles) vs. $\epsilon$, obtained from
Eq. (\ref{state}) with $N=32$. The dashed lines show for
comparison the semiclassical solution of fig.\ref{fig2}. }
\label{fig7}
\end{figure}

From the state (\ref{state}) and the correspondence between $p$
and $\epsilon$ we may evaluate the average populations
$P_i=\langle N_i\rangle/N$, with $i=0,1,2$, as a function of
$\epsilon$. The result is compared in fig.\ref{fig7} with the
semiclassical solution (\ref{sc:1}) and (\ref{sc:2}), for $N=32$.
We observe that the quantum solution has not a sharp transition at
$\epsilon=1/\sqrt{3}$ as the classical one, but there a tail which
becomes negligible for $N\gg 1$.

\subsection{Wigner function}

\begin{figure}
\begin{center}
\includegraphics[width=8.5cm]{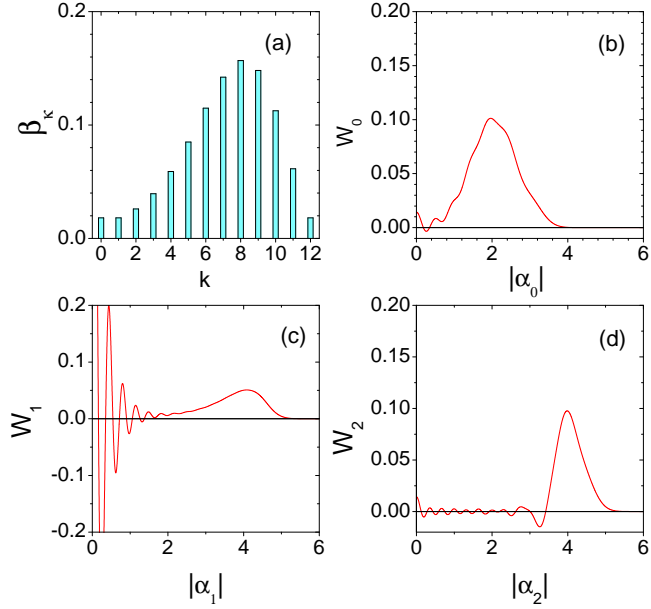}
\end{center}
\vskip 0.2cm \caption{Subradiant state for $N=36$, $p=12$ and
$\epsilon=1/2$: (a) probability $\beta_k$ vs. $k$; (b) $W_0$ vs.
$|\alpha_0|$; (c) $W_1$ vs. $|\alpha_1|$; (d) $W_2$ vs.
$|\alpha_2|$.} \label{ex1}
\end{figure}

Here we show the Wigner function of the subradiance state
$|sr\rangle$ in order to get some more properties of the system.
We start from the definition
\begin{equation}
W(\alpha_{0},\alpha_{1},\alpha_{2})=\int\prod_{i=0}^{2}
\frac{d^{2}\xi_i}{\pi^2}
\;e^{\xi_{i}^{*}\alpha_{i}-\alpha_{i}^{*}\xi_{i}}
\chi(\xi_0,\xi_1,\xi_2)\:, \label{wignerdef}
\end{equation}
where $\alpha_i$ and $\xi_i$ are complex numbers and $\chi$ is the
characteristic function defined as
\begin{eqnarray}
\chi(\xi_0,\xi_1,\xi_2) &=& \langle sr|\hat{D}_0(\xi_0)
\hat{D}_1(\xi_1) \hat{D}_2(\xi_2)|sr\rangle \label{chfun},
\end{eqnarray}
where $\hat{D}_j(\xi_j)=\exp(\xi_j\hat{c}^\dag_j - \xi^{*}_j
\hat{c}_j)$ is a displacement operator for the $j$-th mode. A
straightforward calculation, reported in the Appendix \ref{App2},
yields
\begin{eqnarray}
&&\chi(\xi_0,\xi_1,\xi_2) =
e^{-(|\xi_0|^2+|\xi_1|^2+|\xi_2|^2)/2}\times \nonumber\\
&&\sum_{k=0}^p\beta_k\;L_{p-k}(|\xi_0|^2)\;L_{2k}(|\xi_1|^2)\;L_{N-p-k}(|\xi_2|^2)
\label{chsr},
\end{eqnarray}
and
\begin{eqnarray}
&&W(\alpha_{0},\alpha_{1},\alpha_{2})
=\left(\frac{2}{\pi}\right)^3
e^{-2(|\alpha_0|^2+|\alpha_1|^2+|\alpha_2|^2)}\times \nonumber\\
&&\sum_{k=0}^p\beta_k\;L_{p-k}(4|\alpha_0|^2)\;L_{2k}(4|\alpha_1|^2)\;L_{N-p-k}(4|\alpha_2|^2)
\label{wigsr},
\end{eqnarray}
where
\begin{equation}\label{beta}
    \beta_k=C_p^2\;\frac{(2k)!(N-p-k)!}{(p-k)!(k!)^2}\left(\frac{1}{2\epsilon}\right)^{2k}
\end{equation}
and $L_n(x)$ is the Laguerre polynomial. Notice that the Wigner
function depends only on the modulus of $\alpha_i$ and not from
its phase. As expected, in general it is negative due to the
presence of the Laguerre polynomials. By integrating over the
other two mode variables, from Eq.(\ref{wigsr}) we obtain the
single-mode Wigner functions:
\begin{eqnarray}
  W(\alpha_{0})
&=&\frac{2}{\pi}
e^{-2|\alpha_0|^2}\sum_{k=0}^p(-1)^{p-k}\beta_k\;L_{p-k}(4|\alpha_0|^2)\label{wig1}\\
  W(\alpha_{1}) &=& \frac{2}{\pi}
e^{-2|\alpha_1|^2}\sum_{k=0}^p\beta_k\;L_{2k}(4|\alpha_1|^2)\label{wig2}\\
   W(\alpha_{2}) &=& \frac{2}{\pi}
e^{-2|\alpha_2|^2}\sum_{k=0}^p(-1)^{N-p-k}\beta_k\;\nonumber\\
&& \times L_{N-p-k}(4|\alpha_2|^2).\label{wig3}
\end{eqnarray}
In order to investigate the characteristics of the subradiance
state, let's consider some specific example. An interesting case
is when $\epsilon=1/2$ and $p=N/3$, for which the semiclassical
theory yields $P_1=P_2=4/9$. Fig.\ref{ex1}(a) shows the
probability $\beta_k$ vs. $k$ for $N=36$ and $p=12$. The
probability is maximum for $k=8$, the average value is $\langle
k\rangle=7.14$ and the standard deviation is $\sigma_k=2.64$. The
single-mode Wigner functions $W_i=W(\alpha_i)$ are shown in
fig.\ref{ex1}(b-d): $W_0$ has a maximum at $|\alpha_0|=2$ and
$W_1$ and $W_2$ have a maximum at $|\alpha_1|=|\alpha_2|=4$, in
agreement with the population values predicted by the
semiclassical solution. However, $W_1$ differs considerably from
$W_2$ with a strong oscillation near $|\alpha_1|=0$, probably due
to the tail of the distribution at small $k$ observed in
fig.\ref{ex1}(a). The single-mode Wigner functions present a
pronounced maximum around which they are positive, plus an
oscillating quantum background.

\begin{figure}[b]
\begin{center}
\includegraphics[width=8.5cm]{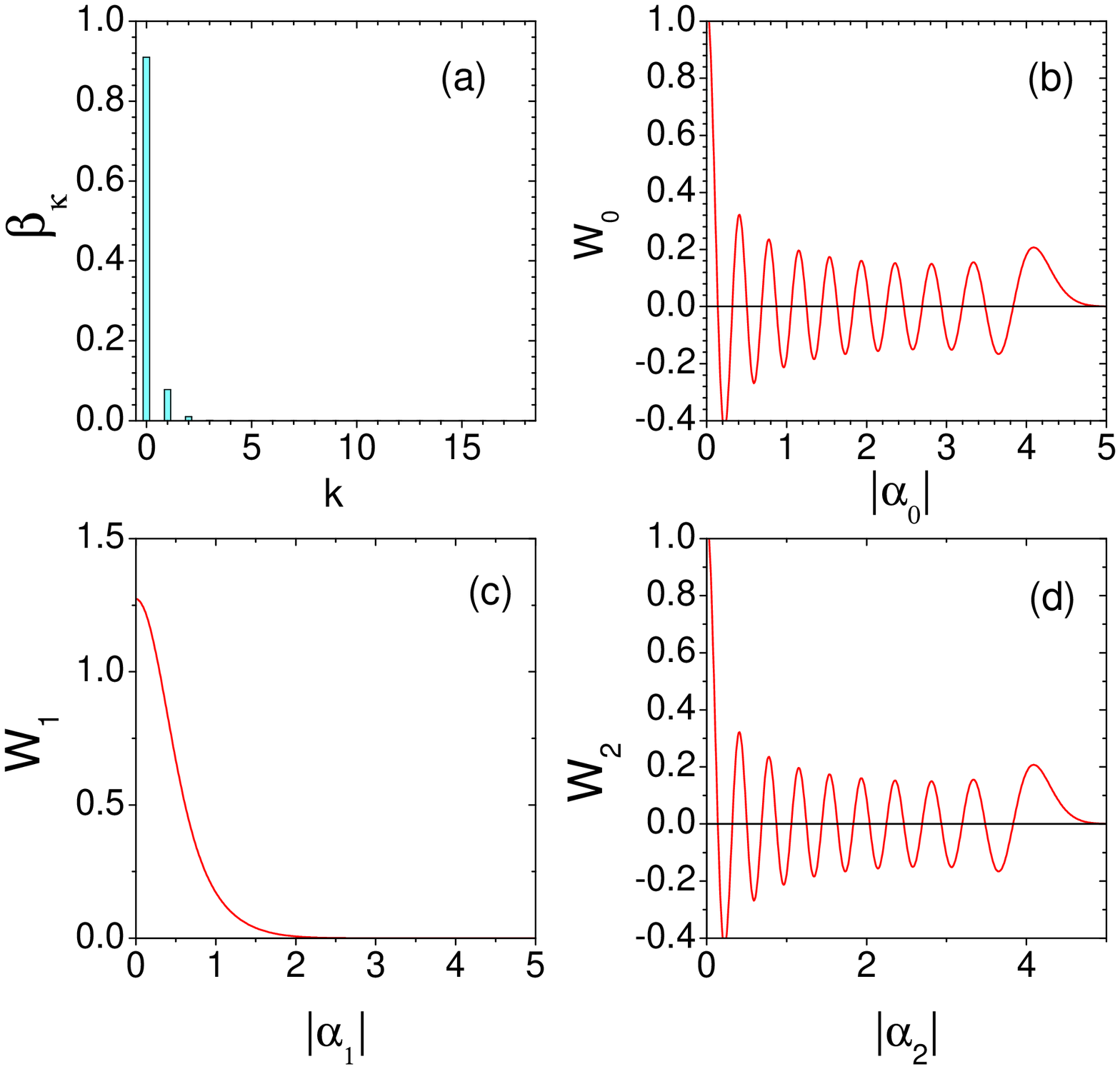}
\end{center}
\vskip 0.2cm \caption{Subradiant state for $N=36$, $p=18$ and
$\epsilon=1+\sqrt{2}$: (a) probability $\beta_k$ vs. $k$; (b)
$W_0$ vs. $|\alpha_0|$; (c) $W_1$ vs. $|\alpha_1|$; (d) $W_2$ vs.
$|\alpha_2|$.} \label{ex2}
\end{figure}

As a second example we consider the case $\epsilon=1+\sqrt{2}$,
for which the semiclassical theory yields $P_1=0$ and
$P_0=P_2=1/2$. In the quantum model it corresponds to the
maximally anti-symmetric state (\ref{state:N/2}) with $p=N/2$.
Fig.\ref{ex2}(a) shows the probability $\beta_k$ vs. $k$ for
$N=36$ and $p=18$. The probability is maximum for $k=0$ and
decreases rapidly to zero for larger $k$, with $\langle
k\rangle=0.1$ and $\sigma_k=0.35$. The single-mode Wigner
functions $W_i$ are shown in fig.\ref{ex2}(b-d). $W_0$ and $W_2$
are equal and very similar to the  Wigner function of the number
state $|N/2\rangle$,
$W(\alpha)=(4/\pi)\exp(-2|\alpha|^2)L_{N/2}(4|\alpha|^2)$
\cite{Walls}. Furthermore, $W_1$ is equal to the vacuum Wigner
function $W(\alpha)=(4/\pi)\exp(-2|\alpha|^2)$. In this case
$N/2$ pairs of atoms with momentum $0$ and $2\hbar q$ are
produced.

\begin{figure}
\begin{center}
\includegraphics[width=8.cm]{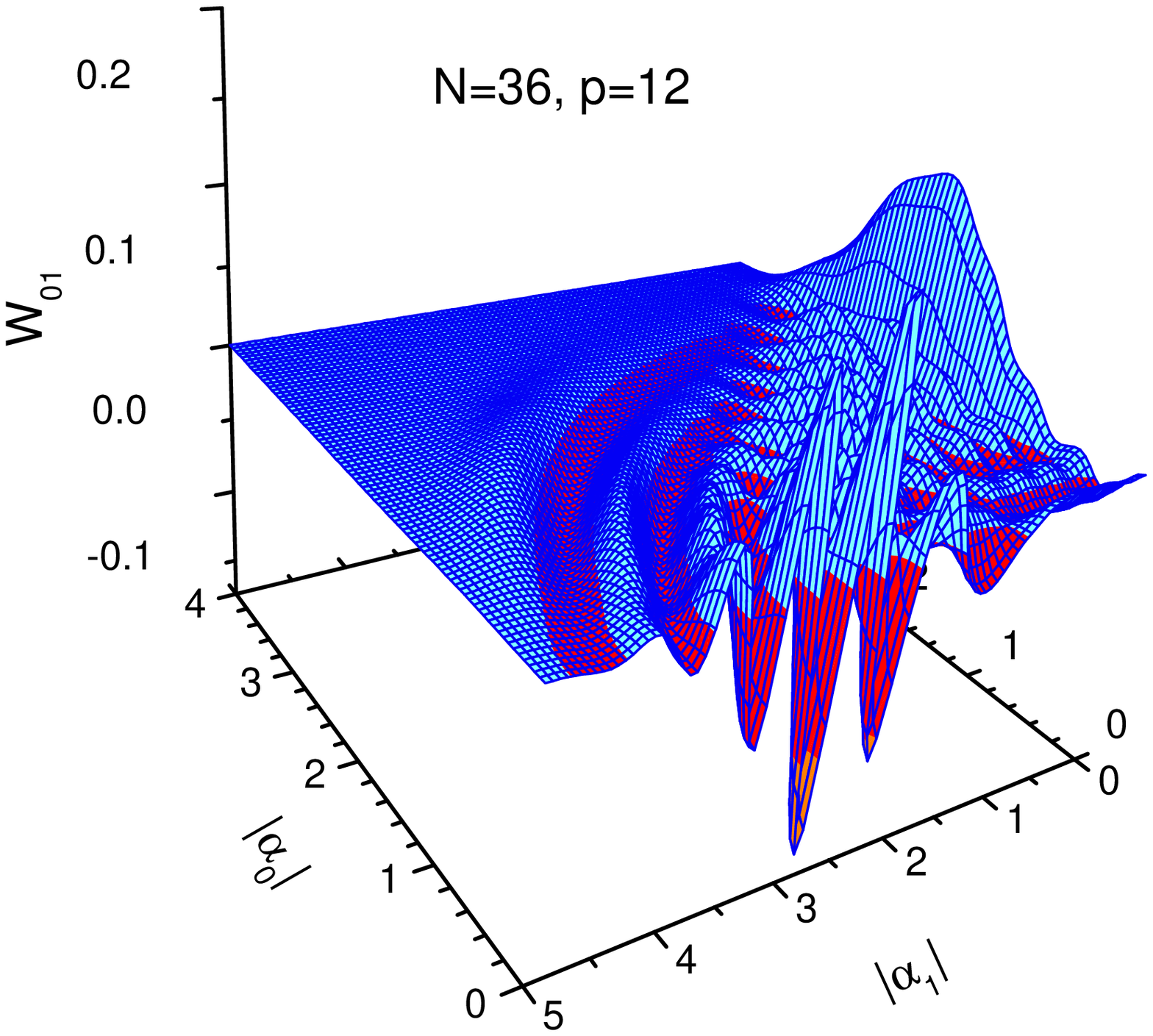}
\end{center}
\vskip 0.2cm \caption{Two-mode Wigner function
$W(\alpha_0,\alpha_1)$  for $N=36$, $p=12$.} \label{figw01}
\end{figure}

\begin{figure}
\begin{center}
\includegraphics[width=8.cm]{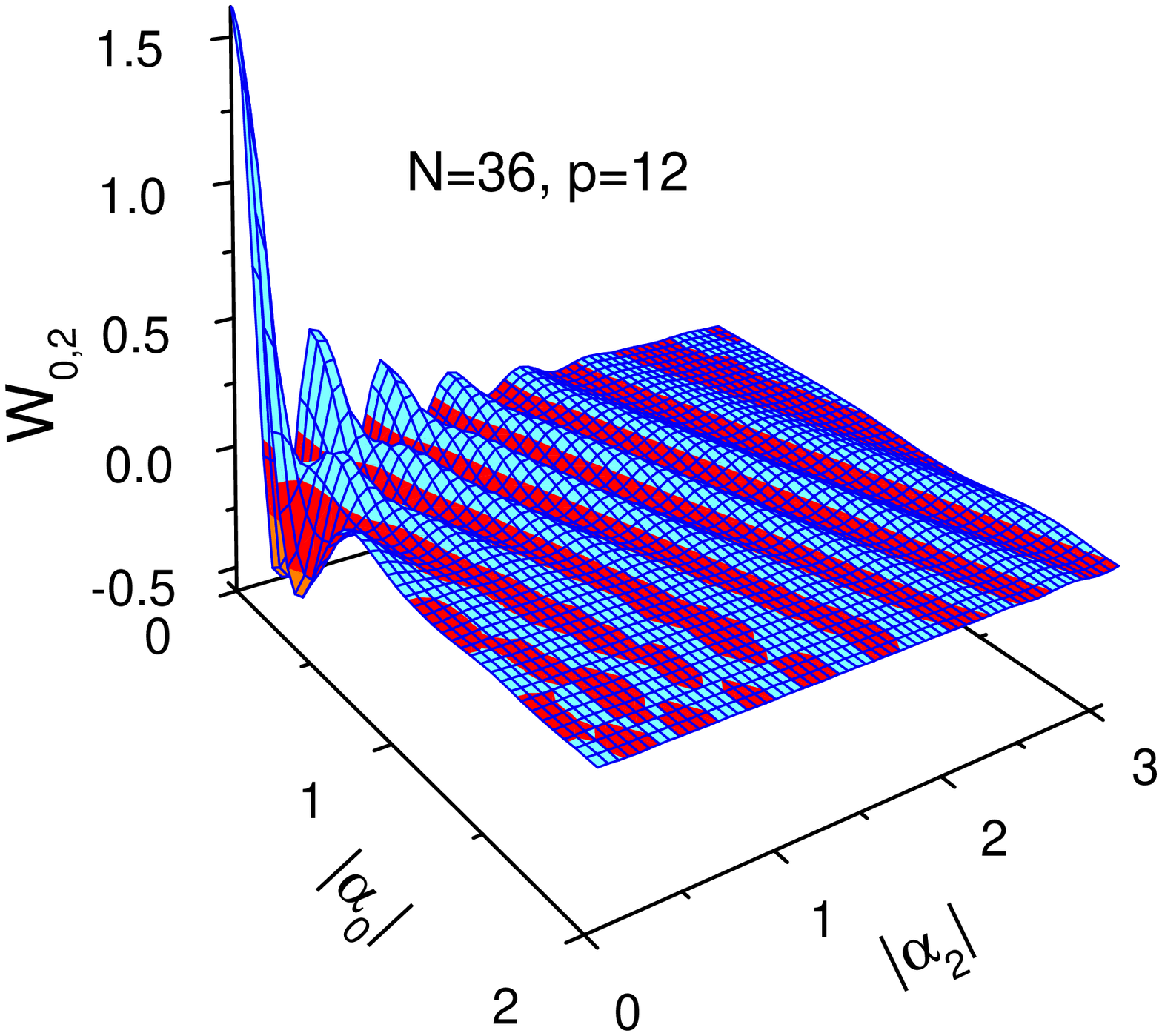}
\end{center}
\vskip 0.2cm \caption{Two-mode Wigner function
$W(\alpha_0,\alpha_2)$  for $N=36$, $p=12$.} \label{figw02}
\end{figure}

\begin{figure}
\begin{center}
\includegraphics[width=8.cm]{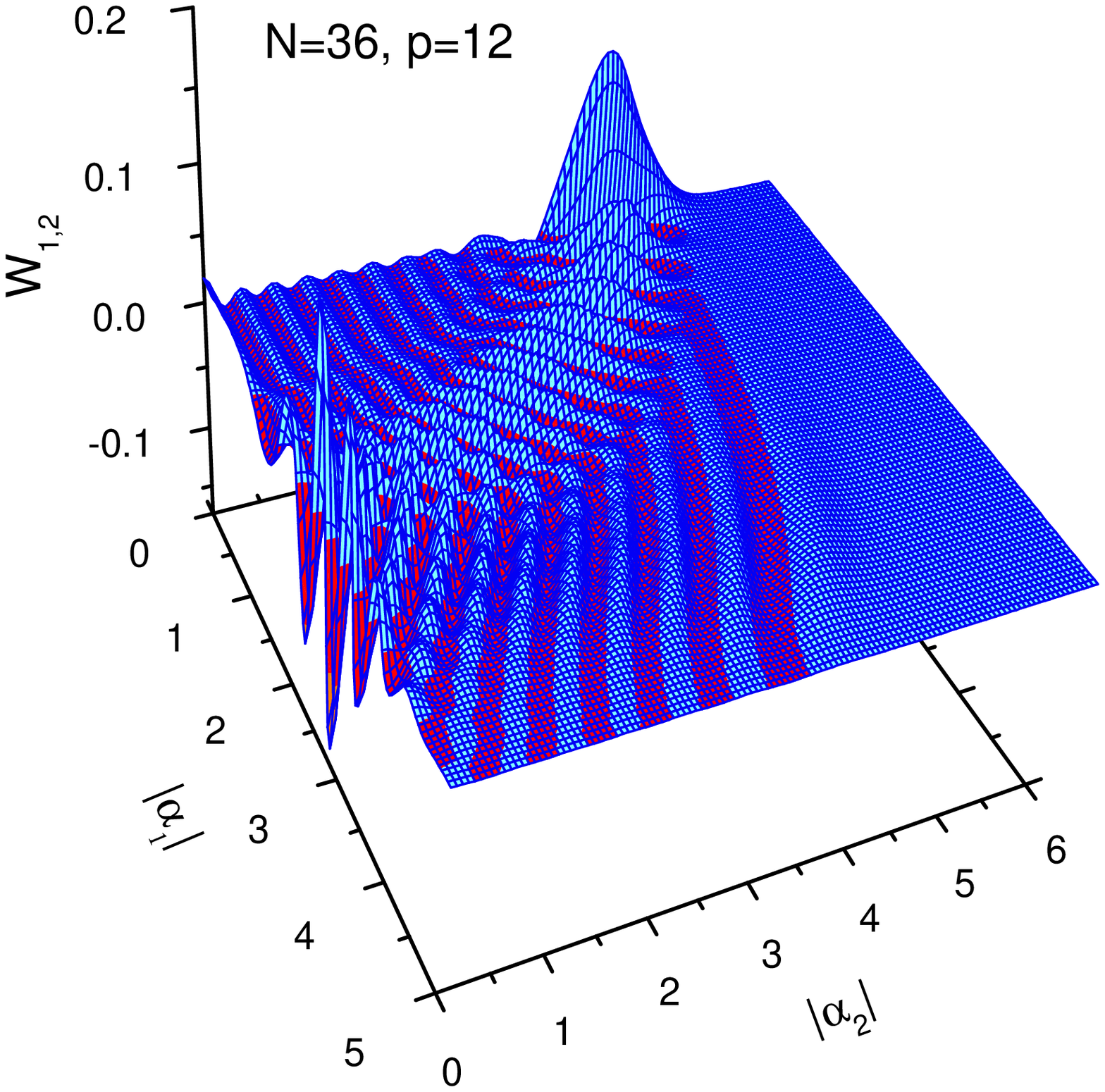}
\end{center}
\vskip 0.2cm \caption{Two-mode Wigner function
$W(\alpha_1,\alpha_2)$  for $N=36$, $p=12$.} \label{figw12}
\end{figure}

\begin{figure}
\begin{center}
\includegraphics[width=8.cm]{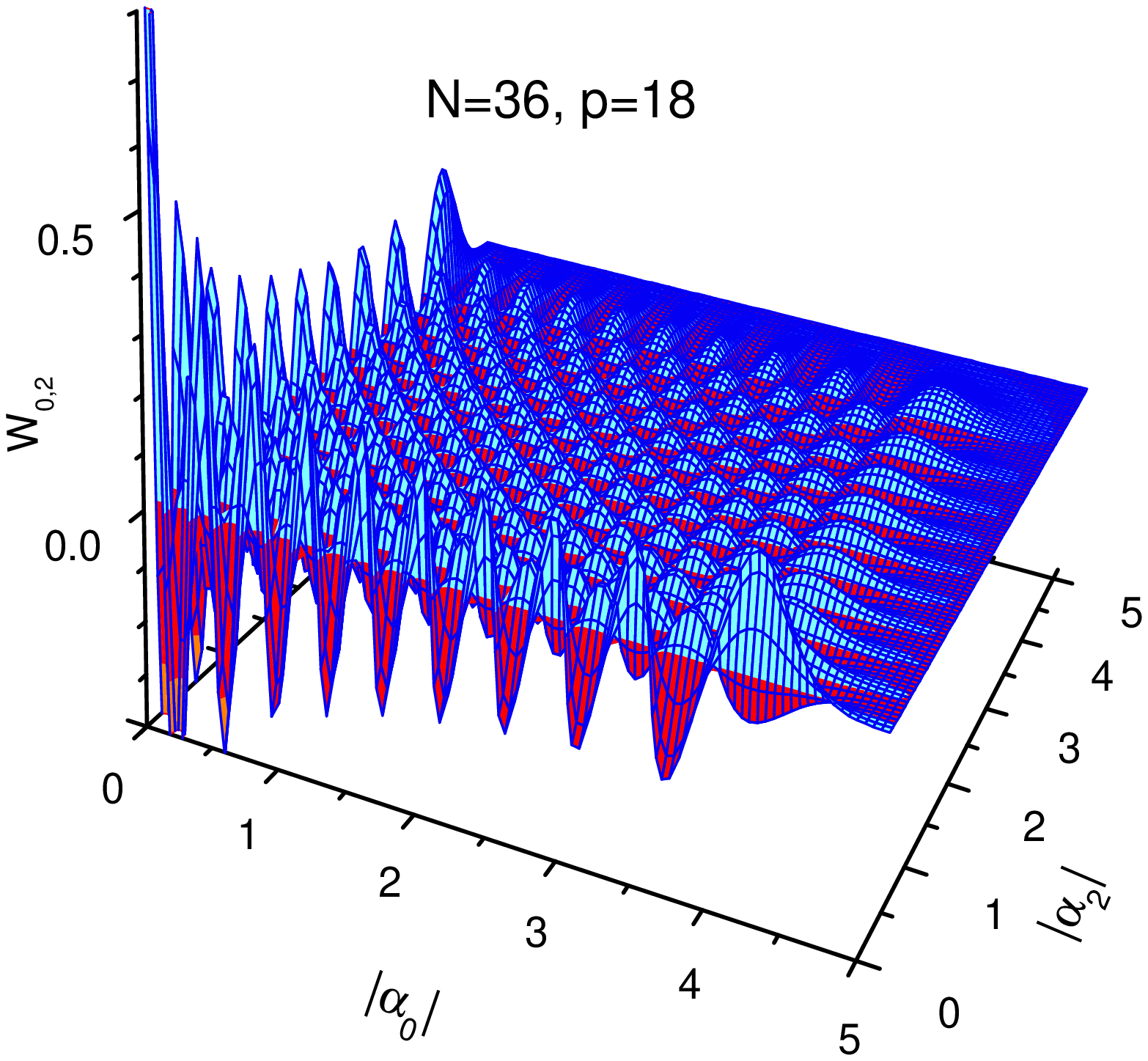}
\end{center}
\vskip 0.2cm \caption{Two-mode Wigner function
$W(\alpha_0,\alpha_2)$  for $N=36$, $p=18$.} \label{figw02a}
\end{figure}

The three-mode Wigner function (\ref{wigsr}), after integrating
one mode variable, yields the following two-mode Wigner functions:
\begin{widetext}
\begin{eqnarray}
  W(\alpha_{0},\alpha_{1})
&=&\left(\frac{2}{\pi}\right)^2 e^{-2(|\alpha_0|^2+|\alpha_1|^2)}
\sum_{k=0}^p(-1)^{p+k}\beta_k L_{p-k}(4|\alpha_0|^2)L_{2k}(4|\alpha_1|^2)\label{wig01}\\
 W(\alpha_{0},\alpha_{2})
&=&\left(\frac{2}{\pi}\right)^2 e^{-2(|\alpha_0|^2+|\alpha_2|^2)}
\sum_{k=0}^p\beta_k L_{p-k}(4|\alpha_0|^2) L_{N-p-k}(4|\alpha_2|^2)\label{wig02}\\
 W(\alpha_{1},\alpha_{2})
&=&\left(\frac{2}{\pi}\right)^2 e^{-2(|\alpha_1|^2+|\alpha_2|^2)}
\sum_{k=0}^p(-1)^{p-k}\beta_k
L_{p-k}(4|\alpha_1|^2)L_{2k}(4|\alpha_2|^2)\label{wig12}
\end{eqnarray}
\end{widetext}
Figures \ref{figw01}, \ref{figw02} and \ref{figw12} (color online)
show the two-mode Wigner functions $W_{i,j}=W(\alpha_i,\alpha_j)$
as a function of $|\alpha_i|$ and $|\alpha_j|$, for $i,j=0,1,2$
for the case $N=36$ and $p=12$, corresponding to $\epsilon=1/2$
(see fig.\ref{ex1}): The red color corresponds to a negative value
of the functions. We observe several zones of negativity
indicating non classical correlations between the two modes. In
particular, $W_{0,1}$ and $W_{1,2}$ show strong correlations of
the modes $0$ and $2$ with the mode $1$, which has strong
oscillations, whereas $W_{0,2}$ is larger near the vacuum value
$(0,0)$. Hence, we can say that qualitatively the modes $0$ and
$2$ behave quasi-classically whereas the mode $1$ has feature
similar to a number state. Figures \ref{figw02a}, shows $W_{0,2}$
vs. $|\alpha_0|$ and $|\alpha_2|$ for the case $N=36$ and $p=18$,
corresponding to $\epsilon=1+\sqrt{2}$ (see fig.\ref{ex2}). The
two-mode Wigner function looks the product of two single-mode
number Wigner functions shown in fig.\ref{ex2}(b) and (d), with a
bi-dimensional regular mesh of positive and negative zones.

\subsection{Atom statistics}

\begin{figure}[b]
\begin{center}
\includegraphics[width=7cm]{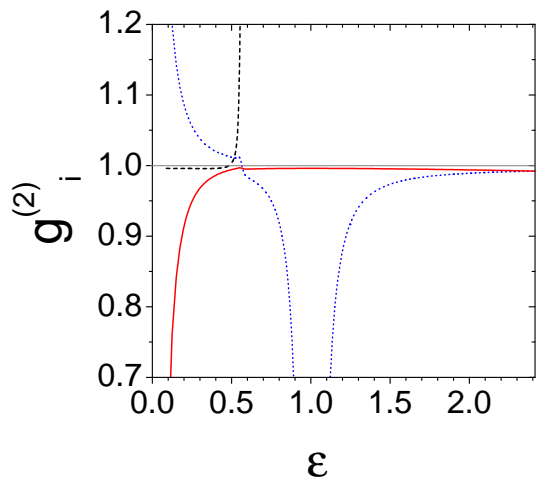}
\end{center}
\vskip 0.2cm \caption{Intensity correlation functions vs.
$\epsilon$: $g_0^{(2)}(0)$ (dotted blue line), $g_1^{(2)}(0)$
(dashed black line) and $g_2^{(2)}(0)$ (continuous red line).}
\label{fig8}
\end{figure}

We calculate now the equal-time intensity correlation and
cross-correlation functions, defined respectively as:
\begin{eqnarray}
g_i^{(2)}(0)&=&\frac{\langle\hat c_i^{\dag}\hat c_i^{\dag}\hat c_i
\hat c_i\rangle}
{\langle\hat N_i\rangle^2}\label{corr1}\\
g_{i,j}^{(2)}(0)&=&\frac{\langle\hat N_i\hat N_j\rangle} {\langle
\hat N_i\rangle\langle\hat N_j\rangle}, \label{corr2}
\end{eqnarray}
with $i=0,1,2$, $i\neq j$ and $\hat N_i=\hat c_i^{\dag}\hat c_i$.
For a classical field there is an upper limit to the second-order
equal-time cross correlation function given by the Cauchy-Schwartz
inequality
\begin{equation}\label{ineq:cl}
g_{i,j}^{(2)}(0)\le[g_i^{(2)}(0)\;g_j^{(2)}(0)]^{1/2}.
\end{equation}
Quantum-mechanical fields, however, can violate this
inequality and are instead constrained by
\begin{equation}
g_{i,j}^{(2)}(0)\le \left[g_i^{(2)}(0)+\frac{1}{\langle \hat
N_i\rangle}\right]^{1/2} \left[g_j^{(2)}(0)+\frac{1}{\langle \hat
N_j\rangle}\right]^{1/2}\label{ineq}
\end{equation}
which reduces to the classical results in the limit of large
occupation numbers. We obtain the following expressions for the
subradiant state:
\begin{eqnarray}
g_0^{(2)}(0)&=&1+\frac{\sigma_k^2-\langle \hat N_0\rangle}{\langle
\hat N_0\rangle^2}\label{g0}\\
g_1^{(2)}(0)&=&1+\frac{4\sigma_k^2-\langle \hat
N_1\rangle}{\langle
\hat N_1\rangle^2}\label{g1}\\
g_2^{(2)}(0)&=&1+\frac{\sigma_k^2-\langle \hat N_2\rangle}{\langle
\hat N_2\rangle^2}\label{g2}
\end{eqnarray}
where $\sigma_k^2=\langle k^2\rangle-\langle k\rangle^2$ and
\[
\langle k^m\rangle=\sum_{k=0}^p k^m\beta_k.
\]
The cross-correlation functions are
\begin{eqnarray}
g_{0,1}^{(2)}(0)&=&1-\frac{2\sigma_k^2}{\langle \hat
N_0\rangle\langle\hat N_1\rangle}\label{g01}
\end{eqnarray}
\begin{eqnarray}
g_{0,2}^{(2)}(0)&=&1+\frac{\sigma_k^2}{\langle \hat
N_0\rangle\langle\hat N_2\rangle}\label{g02}\\
g_{1,2}^{(2)}(0)&=&1-\frac{2\sigma_k^2}{\langle \hat
N_1\rangle\langle\hat N_2\rangle}\label{g12}.
\end{eqnarray}
From (\ref{g0})-(\ref{g12}) it follows that
\begin{widetext}
\begin{eqnarray}
g_{0,1}^{(2)}(0)^2&=&\left(g_0^{(0)}(0)+\frac{1}{\langle \hat
N_0\rangle}\right)\left(g_1^{(2)}(0)+\frac{1}{\langle \hat
N_1\rangle}\right)-\sigma^2\left(\frac{1}{\langle \hat
N_0\rangle}+\frac{2}{\langle \hat N_1\rangle}\right)^2
\label{gg01}\\
g_{0,2}^{(2)}(0)^2&=&\left(g_0^{(0)}(0)+\frac{1}{\langle \hat
N_0\rangle}\right)\left(g_2^{(2)}(0)+\frac{1}{\langle \hat
N_2\rangle}\right)-\sigma^2\left(\frac{1}{\langle \hat
N_0\rangle}-\frac{1}{\langle \hat N_2\rangle}\right)^2
\label{gg02}\\
g_{1,2}^{(2)}(0)^2&=&\left(g_1^{(2)}(0)+\frac{1}{\langle \hat
N_1\rangle}\right)\left(g_2^{(2)}(0)+\frac{1}{\langle \hat
N_2\rangle}\right)-\sigma^2\left(\frac{2}{\langle \hat
N_1\rangle}+\frac{1}{\langle \hat N_2\rangle}\right) \label{gg12}
\end{eqnarray}
\end{widetext}
showing that $g_{ij}^{(2)}(0)$ are consistent with the quantum
inequality (\ref{ineq}).
 Fig.\ref{fig8} shows
the intensity correlation functions $g_i^{(2)}(0)$ for $i=0,1,2$,
for the subradiance state (\ref{state}). We obtain that
$g_2^{(2)}(0)$ (continuous red line) is less than unity for all
the values of $\epsilon$ and $g_0^{(2)}(0)$ (dotted blue line) is
less than unity for $\epsilon> 0.56$. So anti-bunching occurs for
the momentum states $m=0$ and $m=2$, but not for the state $m=1$.

\begin{figure}[t]
\begin{center}
\includegraphics[width=7.5cm]{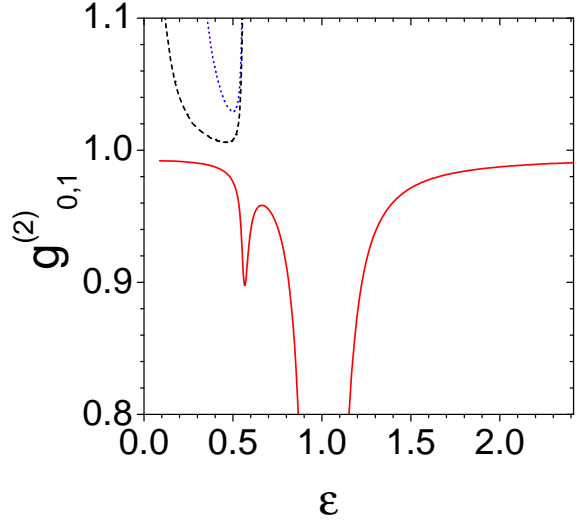}
\end{center}
\vskip 0.2cm \caption{Cross correlation function
$g_{0,1}^{(2)}(0)$ (continuous red line) vs. $\epsilon$. The
dashed black line indicates the classical upper limit of the
inequality (\ref{ineq:cl}); the dotted blue line indicates the
quantum upper limit of the inequality (\ref{ineq}). } \label{fig9}
\end{figure}

\begin{figure}[t]
\begin{center}
\includegraphics[width=8cm]{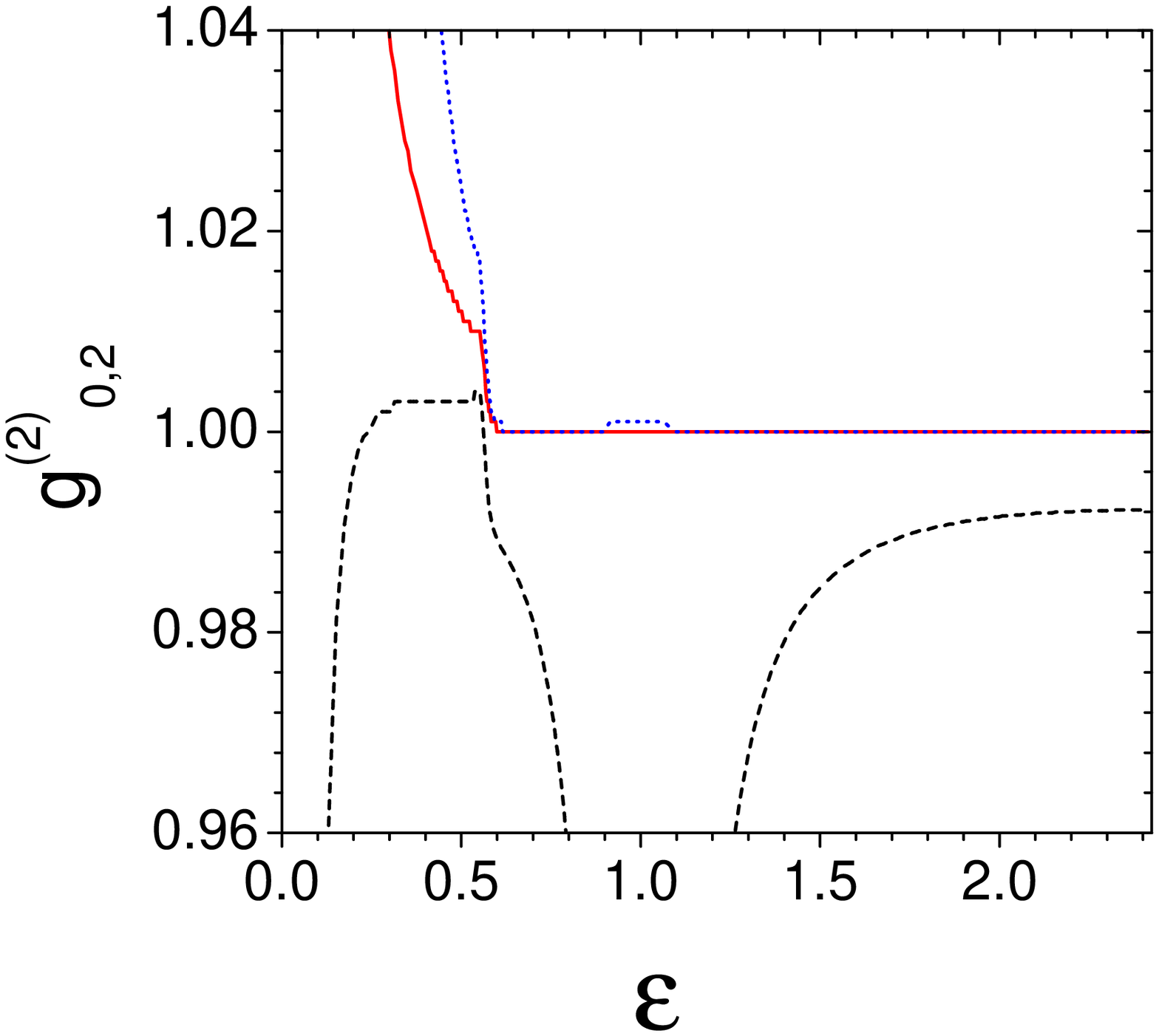}
\end{center}
\vskip 0.2cm \caption{Same as in fig.\ref{fig9} but for
$g_{0,2}^{(2)}(0)$.} \label{fig10}
\end{figure}

\begin{figure}[t]
\begin{center}
\includegraphics[width=8cm]{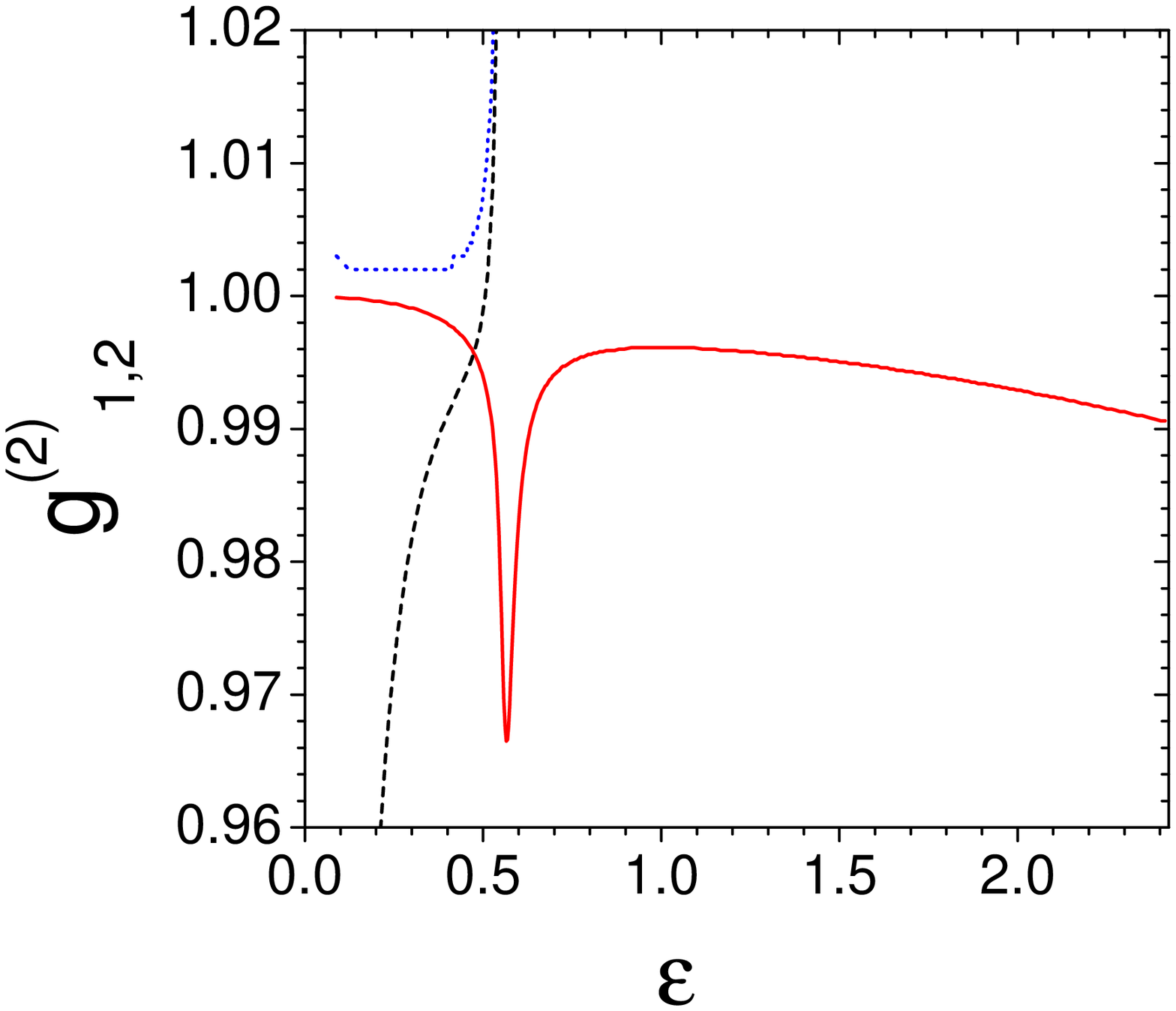}
\end{center}
\vskip 0.2cm \caption{Same as in fig.\ref{fig9} but for
$g_{1,2}^{(2)}(0)$.} \label{fig11}
\end{figure}

Fig.\ref{fig9}-\ref{fig11} show the eventual violation of the
Cauchy-Schwartz inequality (\ref{ineq:cl}) for $g_{i,j}^{(2)}$
(continuous line). The dashed line is the classical upper limit
$[g_{i}^{(2)}(0)g_{j}^{(2)}(0)]^{1/2}$ and the dotted line is the
quantum upper limit $[(g_{i}^{(2)}(0)+1/\langle \hat
N_i\rangle)(g_{j}^{(2)}(0)+1/\langle\hat N_j\rangle)]^{1/2}$. We
found that $g_{0,1}^{(2)}$ is always consistent with the classical
inequality, whereas $g_{0,2}^{(2)}$  and $g_{1,2}^{(2)}$  violate
the Cauchy-Schwartz inequality (\ref{ineq:cl}) for all the
$\epsilon$ and for $\epsilon<0.47$, respectively, showing the
existence of quantum correlations between the modes $m=0$ and
$m=2$ and between the modes $m=1$ and $m=2$. For
$\epsilon>1/\sqrt{3}$, $g_{0,2}^{(2)}$ is close to the upper limit
of the quantum inequality (\ref{ineq}).

\section{conclusions}

We have investigated a possible way to observe subradiance in a
Bose-Einstein condensate in a high-finesse ring cavity, scattering
photons from a two-frequency pump laser into a single-frequency
cavity mode via the quantum collective atomic recoil lasing
(QCARL) mechanism. Subradiance occurs in a degenerate cascade
between three motional levels separated by the two-photon momentum
recoil $\hbar\vec q$, where $\vec q=\vec k-\vec k_s$ is the
momentum transfer between pump and cavity mode. The observation of
subradiance in momentum transitions of cold atomic samples
presents several advantages and differences with respect to the
electronic transitions of excited two-level atoms. First, the
momentum transitions are not affected by the spontaneous emission
if the pump laser is sufficiently detuned from the atomic
resonance. Second, the atomic condensates have a long life and a
very long coherence time, allowing the preparation and the further
manipulation of the subradiant state. Third, the subradiance is
realized among collective motional states containing a large
number of atoms. Subradiance as well superradiance do not need
that the dimension of the sample is smaller than the radiation
wavelength, as in superradiance by excited atoms. For these
reasons, subradiance between motional states of ultracold atoms
may be important for the study of the decoherence-free subspaces
sought in quantum information \cite{dfss}. Other recent proposals
of realizing subradiance in matter wave require a very fine
control of single atoms in optical cavities \cite{foldi}, which
can be very problematic experimentally. On the other hand, the
experimental activity on Superradiant Rayleigh scattering and CARL
with Bose-Einstein condensates \cite{MIT:1,Tub:PRL} has achieved
important progresses and a subradiance experiment with BEC in a
ring cavity could be feasible with the present day techniques. At
ultracold temperature and with a coupling constant $g\sqrt{N}$
much less than the recoil frequency $\omega_r$ should be possible,
using two laser fields with frequency difference $2\omega_r$, to
restrict the momentum transition to only the first two recoil
momentum states. Recent experiments on superradiant scattering
from  a BEC pumped by a two-frequency laser beam
\cite{TFP:1,TFP:2} have shown that the momentum transitions are
enhanced by the presence of the second pump detuned by
$2\omega_r$. However, only inserting the BEC in a high-finesse
ring cavity it will possible to limit the transition sequence to
only two. Then, varying the relative intensity of the two pump
laser beams should be possible to probe the transition from
superradiance to subradiance. As an example of possible
parameters, subradiance could be observed in a ring cavity similar
to that realized in T\"{u}bingen \cite{Tub:PRA} (with length
$L=87$mm, beam waist $w=100\,\mu m$ and finesse $F=5\times 10^5$,
about 5 times the presently achieved value) with
$\kappa=0.2\omega_r$ and $g\sqrt{N}=0.2\omega_r$. This last value
can be obtained using a $^{87}$Rb condensate (with
$\omega_r=(2\pi)15$kHz) with $N=10^{4}$ atoms at a temperature of
some tens of nK, driven by two laser beams with $P_0=174$mW,
$\Delta_0=-(2\pi)3$THz and a  frequency difference precision of
$\delta\omega\le 5$kHz. Subradiance will be reached in about
$100\mu$s, less than the decoherence time of the BEC.

\appendix

\section{The subradiance state\label{App1}}

In order to demonstrate Eq.(\ref{state}) let's consider a state of
the form
\[
|sr\rangle=\sum_{n_{1}=0}^N\sum_{(n_0,n_2)}f(n_0,n_2)|n_0,n_{1},n_2\rangle
\]
where the second sum is over all the pairs $(n_0,n_2)$ such that
$n_0+n_2=N-n_1$. When substituted in (\ref{sr}), it yields
\begin{widetext}
\begin{eqnarray}\label{s:1}
    &&\sum_{n_1=0}^{N-1}
    \sum_{
    \begin{array}{c}
    {_{(n_0,n_2)}}_,{_{n_0\neq 0}}\\
    _{n_0+n_2=N-n_1}
    \end{array}}
    f(n_0,n_2)\sqrt{n_0(n_1+1)}|n_0-1,n_1+1,n_2\rangle\nonumber\\
    &+&
    \epsilon
    \sum_{n_1=1}^{N}
    \sum_{
    \begin{array}{c}
    {_{(n_0,n_2)}}_,{_{n_2\neq N}}\\
    _{n_0+n_2=N-n_1}
    \end{array}}
    f(n_0,n_2)\sqrt{n_1(n_2+1)}|n_0,n_1-1,n_2+1\rangle=0.
\end{eqnarray}
After having redefined the indexes $n_1$ and $n_2$ in the sums, it
becomes
\begin{equation}\label{s:2}
    \left\{
    \sum_{n_1=1}^{N}
    \sum_{
    \begin{array}{c}
    {_{(n_0,n_2)}}\\
    {_{n_0\neq N}}
    \end{array}}
    f(n_0+1,n_2)\sqrt{(n_0+1)n_1}
    +\epsilon
    \sum_{n_1=0}^{N-1}
    \sum_{
    \begin{array}{c}
    {_{(n_0,n_2)}}\\
    {_{n_2\neq 0}}
    \end{array}}
    f(n_0,n_2-1)\sqrt{(n_1+1)n_2}
    \right\}|n_0,n_1,n_2\rangle=0.
\end{equation}
The terms in the curl bracket of Eq. (\ref{s:2})  vanish when the
following conditions are met:
\begin{enumerate}
    \item In the second sum on $n_1$ in the curl bracket the term with $n_1=N$
    is missing: since in the first sum $n_0=n_2=0$ when $n_1=N$, it yields $f(1,0)=0$.
    \item In the first sum on $n_1$ in the curl bracket the term with $n_1=0$ is missing:
    then the second sum on $n_1$ with $n_1=0$, $n_2=N-n_0$ and $n_2\neq 0$
    yields
    \begin{equation}\label{s:4}
    f(n_0,N-n_0-1)=0\quad,\quad n_0=0\dots,N-1.
    \end{equation}
    \item The remaining sum in (\ref{s:2}) yields:
    \begin{equation}\label{s:5}
    \sum_{
    \begin{array}{c}
    {_{(n_0,n_2)}}\\
    _{n_0+n_2=N-n_1}
    \end{array}}
    f(n_0+1,n_2)\sqrt{(n_0+1)n_1}
    +\epsilon
    \sum_{
    \begin{array}{c}
    {_{(n_0,n_2)}}_,{_{n_2\neq 0}}\\
    _{n_0+n_2=N-n_1}
    \end{array}}
    f(n_0,n_2-1)\sqrt{(n_1+1)n_2}=0,
\end{equation}
with $n_1=1,\dots,N-1$. In the second sum of Eq.(\ref{s:2}) the
term $n_2=0$ is missing, so in the first sum the term with $n_2=0$
and $n_0=N-n_1$ vanishes
 and yields $f(k,0)=0$ with $k=2,\dots,N$. Together with the
result of item 1, it yields
\begin{equation}\label{s:6}
    f(k,0)=0\quad,\qquad k=1,2,\dots,N
    \end{equation}
whereas the remaining terms of Eq.(\ref{s:5}) yield
    \begin{equation}\label{s:7}
    \sqrt{(n_0+1)n_1} f(n_0+1,N-n_0-n_1)
    =-\epsilon
    \sqrt{(n_1+1)(N-n_0-n_1)}f(n_0,N-n_0-n_1-1)
\end{equation}
with $n_1=1,\dots,N-1$ and $n_0=0,\dots,N-n_1$.
\end{enumerate}
For $n_1=2$ Eq.(\ref{s:7}) yields
    \begin{equation}\label{s:8}
    \sqrt{2(n_0+1)}f(n_0+1,N-n_0-2)
    =-\epsilon
    \sqrt{3(N-n_0-3)}f(n_0,N-n_0-3)
    \quad,\quad
    n_0=0,\dots,N-3
\end{equation}
Since from Eq.(\ref{s:4}) $f(n_0+1,N-n_0-2)=0$, then
   \begin{equation}\label{s:9}
    f(n_0,N-n_0-3)=0\quad,\quad n_0=0\dots,N-3.
    \end{equation}
Continuing with all the even values of $n_1$, it is easy to show
that
   \begin{equation}\label{s:10}
    f(n_0,N-n_0-k)=0\quad,\quad k=1,3,\dots,N-n_0\,(odd), \quad n_0=0\dots,N-k.
    \end{equation}
 Hence, the only  terms different from zero are those with $n_1=2q+1$ and
 $q=0,\dots,N/2-1$, yielding:
    \begin{equation}\label{s:11}
   f(n_0+1,N-n_0-2q-1)
    =-\epsilon
    \sqrt{\frac{(2q+2)(N-n_0-2q-1)}{(n_0+1)(2q+1)}}f(n_0,N-n_0-2q-2),
\end{equation}
where $n_0=0,\dots,N-(2q+1)$. Eq.(\ref{s:11}) provides a
recurrence relation for the index $n_0$ with a given $n_0+q$. In
fact the difference between the first and second index of $f(a,b)$
in both the left and right terms of Eq.(\ref{s:11}) is
$\Delta=b-a=N-2(n_0+q+1)$. So, introducing the new index
$p=n_0+q+1$, Eq.(\ref{s:11}) can be written, for $p=1,\dots,N/2$,
as:
   \begin{equation}\label{s:12}
   f(p-q-1,N-p-q-1)
    =-\frac{1}{\epsilon}
    \sqrt{\frac{(2q+1)(p-q)}{(2q+2)(N-p-q)}}f(p-q,N-p-q)
    \quad,\quad
    q=0,\dots,p
\end{equation}
By iteration of Eq.(\ref{s:12}) we obtain
   \begin{equation}\label{s:13}
   f(p-q,N-p-q)
    =\left(-\frac{1}{\epsilon}\right)^q
    \sqrt{\frac{(2q+1)!!}{(2q)!!}\frac{p!(N-p-q)!}{(p-q)!(N-p)!}}f(p,N-p)
    \quad,\quad
    q=0,\dots,p
\end{equation}
where $(2k)!!=2k\cdot(2k-2)\dots 2\cdot 1$ and
$(2k+1)!!=(2k+1)\cdot(2k-1)\dots 3\cdot 1$. Since
$(2q+1)!!/(2q)!!=(2q)!/(2^qq!)^2$, finally we obtain :
\begin{equation}\label{s:state}
|sr\rangle=C_p\sum_{q=0}^p \left(-\frac{1}{2\epsilon}\right)^q
    \sqrt{\frac{(2q)!(N-p-q)!}{(q!)^2(p-q)!}}
    |p-q,2q,N-p-q\rangle
\end{equation}
where $C_p=[p!/(N-p)!]^{1/2}f(p,N-p)$.
\end{widetext}

\section{Derivation of the subradiant Wigner function\label{App2}}

We demonstrate Eq.(\ref{chsr}). Writing the displacement operator
as a product of operators,
$\hat{D}(\xi)=\exp(-|\xi|^2/2)\exp(\xi\hat{c}^\dag) \exp(-\xi^{*}
\hat{c})$ and using the formula
\[
\exp(-\xi^{*}
\hat{c})|k\rangle=\sum_{n=0}^k\frac{(-\xi^*)^n}{n!}\sqrt{\frac{k!}{(k-n)!}}
|k-n\rangle
\]
we obtain
\begin{equation}\label{dis}
\langle k'|\hat{D}(\xi) |k\rangle= e^{-|\xi|^2/2}
L_{k}(|\xi|^2)\delta_{k,k'}
\end{equation}
where
\[
L_{k}(x)=\sum_{n=0}^k (-1)^n \begin{pmatrix}
  k \\
  n \\
\end{pmatrix}\frac{x^n}{n!}
\]
is the Laguerre Polynomial of order $k$. Using (\ref{dis}) in
(\ref{chfun}) with the subradiant state (\ref{state}) we obtain
\begin{eqnarray}
&&\chi(\xi_0,\xi_1,\xi_2) =
e^{-(|\xi_0|^2+|\xi_1|^2+|\xi_2|^2)/2}\times \nonumber\\
&&\sum_{k=0}^p\beta_k\;L_{p-k}(|\xi_0|^2)\;L_{2k}(|\xi_1|^2)\;L_{N-p-k}(|\xi_2|^2)
\label{char}.
\end{eqnarray}
In order to evaluate the Wigner function (\ref{wignerdef}) we must
calculate an integral of the form:
\begin{equation}
I_m(\alpha)=\int d^{2}\xi
\;e^{\xi^{*}\alpha-\alpha^{*}\xi-|\xi|^2/2}L_{m}(|\xi|^2).
\end{equation}
Introducing polar coordinates $\xi=-ir\exp(i\phi)$ and
$\alpha=|\alpha|\exp(i\psi)$ it transforms into
\begin{eqnarray}
&&I_m(\alpha)=\int_0^{\infty}dr\; r
e^{-r^2/2}L_m(r^2)\int_{0}^{2\pi}d\phi
e^{2ir|\alpha|\cos(\phi-\psi)}=\nonumber\\
&&2\pi\int_0^{\infty}dr\; r e^{-r^2/2}L_m(r^2)J_0(2r|\alpha|)
\end{eqnarray}
Where $J_0(x)$ is the Bessel function of zero order. Using the
formula \cite{Grad}:
\begin{eqnarray}
&&\int_0^{\infty}dx\;x
e^{-ax^2/2}L_n(bx^2/2)J_0(xy)=\nonumber\\
&&\frac{(a-b)^n}{a^{n+1}}e^{-y^2/2a}L_n
\left(\frac{by^2}{2a(b-a)}\right)
\end{eqnarray}
we obtain
\begin{equation}\label{II}
I_m(\alpha)=2\pi(-1)^m e^{-2|\alpha|^2}L_m(4|\alpha|^2)
\end{equation}
So, from the definition of Wigner function (\ref{wignerdef}) and
using Eqs. (\ref{char}) and (\ref{II}) we obtain
\begin{equation}\label{wigner}
W(\alpha_{0},\alpha_{1},\alpha_{2})=\frac{1}{\pi^6}\sum_{k=0}^p\beta_k\;I_{p-k}(\alpha_0)
I_{2k}(\alpha_1)I_{N-p-k}(\alpha_2)
\end{equation}
which coincides with Eq.(\ref{wigsr}). Using the formula
\cite{Grad}
\begin{equation}\label{LL}
\int_0^{\infty}dx \;e^{-x/a}L_m(x)=a(1-a)^{n}
\end{equation}
we have
\begin{equation}\label{norma}
\frac{1}{\pi^2}\int d^{2}\alpha\;I_m(\alpha)=1.
\end{equation}
From (\ref{II}), (\ref{wigner}) and (\ref{norma}) we obtain the
expressions (\ref{wig1})-(\ref{wig3}) and
(\ref{wig01})-(\ref{wig12}) of the one-mode and two-mode reduced
Wigner functions, respectively.

\end{document}